\documentstyle[12pt]{article}
\begin{document}

%
\begin{titlepage}

\title{Bound $q\bar{q}$ systems in the framework of two-body Dirac
equations obtained from different versions of 3D-reductions of the
Bethe-Salpeter equations}

\author{T.Babutsidze, T.Kopaleishvili, and D.Kurashvili 
\\
{\footnotesize High Energy Physics Institute, Tbilisi
State
University 380086 University Str. 9, Tbilisi, Georgia}
}

\maketitle

\abstract

The two-body Dirac equations for the bound $q\bar{q}$ systems are
obtained from the different (five) versions of the 3D-equations
derived from Bethe-Salpeter equation with the instantaneous kernel
in the momentum space using the additional approximations. There
are formulated the normalization conditions for the wave functions
satisfying the obtained two-body Dirac equations. The spin
structure of the confining $qq$ interaction potential is taken in
the form $x\gamma^{0}_{1}\otimes\gamma^{0}_{2}+(1-x)I_{1} \otimes
I_{2}$, with $0 \leq x \leq 1$. It is shown that the two-body
Dirac equations obtained from the Salpeter equation does not
depend on $x$. As to other four versions such a dependence is
left. For the systems ($u\bar{s}$), ($c\bar{u}$), ($c\bar{s}$) the
dependence of the stable solutions of the Dirac equations obtained
in the different version on the mixture parameter $x$ is
investigated and results is compared with such dependence of
3D-equations derived from Bethe-Salpeter equations without the
additional approximation and some new conclusions are obtained.

\begin{center}

\end{center}

\end{titlepage}

%
%
\section{Introduction}

The Bethe-Salpeter (BS) equation provides a natural basis for the
relativistic treatment of bound $q\bar{q}$ systems in the
framework of the constituent quark model. But due to fact that the
BS wave function (amplitude) has not probability interpretation,
three-dimensional (3D) reduction is necessary. Review of
investigations of bound $q\bar{q}$ systems (mesons) on basis of
equations obtained in different versions of 3D-reduction of BS
equation in the instantaneous (static) approximation for kernel of
BS equation is given in Ref. \cite{b1}. In literature there are
known five such versions formulated in Refs. \cite{b1}-\cite{b7},
below noted as SAL \cite{b2}, GR \cite{b3}, MW \cite{b4}, CJ
\cite{b5}and MNK\cite{b6}, \cite{b7} versions. The last four
3D-equations have correct one-body limit (the Dirac equation) when
the mass of one of the particles tends to infinity. As it is
well-known the Salpeter has not such a limit. Note that Gross
equation is obtained only for $m_1 \neq m_2$ case, while other
versions work for the equal masses ($m_1 = m_2$) too.

Below we shall consider the problem how to get two-body Dirac
equations for the bound $q\bar{q}$ systems from above mentioned
3D-relativistic equations, and how to formulate corresponding
normalization conditions for the wave functions. Then, these
equations will be used for the investigation of some aspects of the
problem connected to the mass spectra for bound $q\bar{q}$ systems
(mesons). Namely, the dependence of the existence of stable solutions 
of obtained equations and of the mass spectra on the Lorentz(spin) 
structure of confinement $q\bar q$ interaction potential will be studied.
Further, the comparison to the results obtained without
additional approximations will be made.

%
%
\section{The two-body Dirac equation for bound $q\bar{q}$ systems
and normalization conditions for the corresponding wave functions}

To derive such an equation note that all 3D-equations given in
Ref. \cite{b1} can be written in the common form (c.m.f.)

\begin{eqnarray}
\label{eq1} [ M - h_1(\textbf{p}) - h_2(-\textbf{p}) ]
\widetilde{\Phi}_M (\textbf{p}) = \Pi(M; \textbf{p}) \gamma_1^0
\otimes \gamma_2^0 \int \frac{d^3 \textbf{p}'}{(2\pi)^3}{V}
(\textbf{p}, \textbf{p}')\widetilde{\Phi}_M (\textbf{p}')
\end{eqnarray} where

\begin{eqnarray}\label{eq2}
\Pi(M; \textbf{p}) =
\begin{array}{l l}
\frac{1}{2} \left( \frac{h_1}{\omega_1} +
\frac{h_2}{\omega_2}\right), & \mbox{(SAL)} \\
\frac{1}{2} \left( 1 + \frac{h_1}{\omega_1}\right), & \mbox{(GR)} \\
\frac{1}{2} \left( \frac{h_1}{\omega_1} +
\frac{h_2}{\omega_2}\right) + \frac{M}{\omega_1 + \omega_2}\left(
1 - \frac{h_1}{\omega_1} \otimes \frac{h_2}{\omega_2} \right) ,
& \mbox{(MW)}\\
\frac{1}{2} \left[ \frac{M+h_1+h_2}{\omega_1+\omega_2}  \right],
& \mbox{(CJ)} \\
\frac{1}{2} \left[ \frac{M( a - (p^{+}_{0})^2) - p^{+}_{0} M
(h_1-h_2) + B(h_1+h_2) }{BR} \right], & \mbox{(MNK)}
\end{array} \\
a = E_1^2 - E_2^2 = \frac{1}{4} \left( M^2 + b_0^2 -2(\omega_1^2 +
\omega_2^2 ) \right), \;  b_0 = E_1 - E_2, \; E_{(^{1}_{2})} =
\frac{M}{2} \left( 1 \pm d_{12} \right) \nonumber \\
d_{12} = \frac{m_1^2-m_2^2}{M^2}, \; p^{+}_0 = \frac{R-b}{2y}, \;
y = \frac{m_1 - m_2}{m_1 + m_2}, \; R = \sqrt{b^2-4y^2a}, \; b =
M+b_0y, \; \nonumber \\
B = a+p^{+}_0 b_0 + (p^{+}_0)^2 \nonumber \\
h_i = \vec{\alpha}_i \vec{p}_i + m_i \gamma_i^0, \; \omega_i =
\sqrt{m_i^2+{\bf{p}}^{2}_{i}}, \; O_1 = O_1\otimes I_2, \; O_2 =
I_1 \otimes O_2 \nonumber
\end{eqnarray}

Note that from the eq.(\ref{eq1}) with the operators $\Pi$
(\ref{eq2}) immediately follow the system of equations (3.61) in
Ref.\cite{b1} with definition (3.61-63), if eq.(\ref{eq1}) is
multiplied from left by projection operator
$\Lambda^{(\alpha_1\alpha_2)}_{12}$ and are used their properties:

\begin{eqnarray}
\label{eq3}
\begin{array}{l}\Lambda^{(\alpha_1\alpha_2)}_{12} =
\Lambda^{(\alpha_1)}_{1} \otimes \Lambda^{(\alpha_2)}_{2}, \, \,
\, \Lambda^{(\alpha_i)}_{i} = \frac{\omega_i+\alpha_i
h_i}{2\omega_i}, \,\,\,
\Lambda^{(\alpha_i)}_{i}\Lambda^{(\beta_i)}_{i}=\delta_{\alpha_i\beta_i}\Lambda^{(\alpha_i)}
\\
\\
\Pi^{SAL} =
\Lambda_{12}^{(++)}-\Lambda_{12}^{(--)}=\frac{1}{2}\left(
\frac{h_1}{\omega_1} +  \frac{h_2}{\omega_2} \right),\,\,\,
\Pi^{GR}=\Lambda_{12}^{(++)}+\Lambda_{12}^{(+-)}=\frac{1}{2}\left(
1 +  \frac{h_1}{\omega_1} \right)
\end{array}
\end{eqnarray}

Now if in the operator $\Pi^{SAL}$ use the approximation

\begin{eqnarray}
\label{eq4}
\frac{h_i}{\omega_i} \Longrightarrow
\frac{h_i}{\omega_i} \mid^{\bf{p}_i \rightarrow 0 } = \gamma_i^0
\end{eqnarray} then we obtain the two-body Dirac equation

\begin{eqnarray}
\label{eq5} [ M - h_1( {\bf{p}} ) - h_2( - {\bf{p}} ) ] \Psi_M (
{\bf{p}} ) = \Pi_0^{SAL} \gamma_1^0 \otimes \gamma_2^0 \int
\frac{d^3 \bf{p}'}{(2\pi)^3}{V} ({\bf{p}}, {\bf{p}}') \Psi_M
(\bf{p}')
\end{eqnarray}

\begin{eqnarray}
\label{eq5b} \Pi_0^{SAL} = \frac{1}{2}\left( \gamma_1^0 +
\gamma_2^0 \right)
\end{eqnarray} which already was used for bound $q\bar{q}$ systems in
Refs.\cite{b8},\cite{b9}.

In approximation (\ref{eq4}) from (\ref{eq2}) follows

\begin{eqnarray}
\label{eq6}
\Pi_0^{GR} = \frac{1}{2}\left( 1+\gamma_1^0 \right)
\end{eqnarray}.

As to MW, CJ and MNK versions for derivation of corresponding
two-body Dirac equations the additional to (\ref{eq4})
approximation is need, namely

\begin{eqnarray}
\label{eq4b} \Pi(M;0) \Longrightarrow \Pi(m_1+m_2;0) \equiv \Pi_0
\end{eqnarray}
which is quite natural because it corresponds to zero
approximation in iteration procedure for solving nonlinear over
$M$ eq.(\ref{eq1}) for the MW, CJ and MNK versions. As a result
from (\ref{eq2}) can be obtained

\begin{eqnarray}
\label{eq6b}
\Pi_0^{MW} = \frac{1}{2} \left[ \gamma_1^0+\gamma_2^0
+ (1-\gamma_1^0\otimes\gamma_2^0) \right]
\end{eqnarray}

\begin{eqnarray}
\label{eq6c} \Pi_0 = \frac{1}{2} \left[ 1 + \mu_1
\gamma_1^0+\mu_2\gamma_2^0 \right],
\begin{array}{l l}
\mu_i = \frac{m_i}{m_1+m_2} & \mbox{(CJ)} \\
\\
\mu_i = \frac{m_i^2}{m_1^2+m_2^2} & \mbox{(MNK)} \\
\end{array}
\end{eqnarray}

Thus, we have the following two-body Dirac equations obtained from
(\ref{eq1}), (\ref{eq2})

\begin{eqnarray}
\label{eq7} [ M - h_1( {\bf{p}} ) - h_2( - {\bf{p}} ) ] \Psi_M (
{\bf{p}} ) = \Pi_0 \gamma_1^0 \otimes \gamma_2^0 \int \frac{d^3
\bf{p}'}{(2\pi)^3}{V} ({\bf{p}}, {\bf{p}}') \Psi_M (\bf{p}')
\end{eqnarray} where the operator $\Pi_0$ is given by the formulae
(\ref{eq5b},\ref{eq6},\ref{eq6b},\ref{eq6c}).

Note that there is the another approach for formulation the
two-body Dirac equations, namely, generation of the one-body Dirac
equation to two-body one, using constrain dynamics and relation to
quantum filed theory. Review of such an approach is given in
Ref.\cite{b10}.

Representing the wave function $\Psi_M(\bf{p})$ as sum of
"frequency" components

\begin{eqnarray}
\label{eq8}
\Psi_M({\bf{p}}) = \sum_{\alpha_1\alpha_2}
\Lambda_{12}^{(\alpha_1\alpha_2)}({\bf{p}})\Psi_M({\bf{p}}) =
\sum_{\alpha_1\alpha_2} \Psi_M^{(\alpha_1\alpha_2)}({\bf{p}})
\end{eqnarray} from the eq.(\ref{eq7}) follows the system of the
equation for the functions $\Psi_M^{(\alpha_1\alpha_2)}$

\begin{eqnarray}
\label{eq9}
[ M - (\alpha_1\omega_1 + \alpha_2 \omega_2) ]
\Psi_M^{(\alpha_1\alpha_2)} ( {\bf{p}} ) =
\Lambda_{12}^{(\alpha_1\alpha_2)} \Pi_0 \gamma_1^0 \otimes
\gamma_2^0 \int \frac{d^3 \bf{p}'}{(2\pi)^3}{V} ({\bf{p}},
{\bf{p}}')\sum_{\alpha_1'\alpha_2'} \Psi_M^{(\alpha_1'\alpha_2')} (\bf{p}').
\end{eqnarray}

Taking the $q\bar{q}$ interaction operator ${V}$ in the form
\cite{b1} (combination of one-gluon exchange and confining part of
potential)

\begin{eqnarray}
\label{eq10}
V = \gamma_1^0 \otimes \gamma_2^0 V_{OG}+\left[ x
\gamma_1^0 \otimes \gamma_2^0 + (1-x) I_1 \otimes I_2 \right]V_C
\end{eqnarray} and representing the function
$\Psi_M^{(\alpha_1\alpha_2)}$ as

\begin{eqnarray}
\label{eq11} \Psi_M^{(\alpha_1\alpha_2)}({\mathbf{p}}) =
N_{12}^{(\alpha_1\alpha_2)}(p) \left[ \left( \begin{array}{c} 1 \\
\frac{\alpha_1 \vec{\sigma}_1 \vec{p}}{\omega_1+\alpha_1 m_1}
\end{array} \right) \otimes  \left( \begin{array}{c} 1 \\
\frac{- \alpha_2 \vec{\sigma}_2 \vec{p}}{\omega_2+\alpha_2 m_2}
\end{array} \right) \equiv f^{(\alpha_1 \alpha_2)}_{12}(\vec{p})
\right]\chi_{M}^{(\alpha_1\alpha_2)}(\mathbf{p})
\end{eqnarray} where

\begin{eqnarray}
\label{eq12}
N_{12}^{(\alpha_1\alpha_2)} =
\sqrt{\frac{\omega_1+\alpha_1
m_1}{2\omega_1}}\sqrt{\frac{\omega_2+\alpha_2 m_2}{2\omega_2}}.
\end{eqnarray} then for the wave functions
$\chi_{M}^{(\alpha_1\alpha_2)}$ from (\ref{eq9}) can be obtained
the following system of equations

\begin{eqnarray}
\label{eq13} \left[ M - ( \alpha_1 \omega_1 + \alpha_2 \omega_2 )
\right] \chi_{M}^{(\alpha_1\alpha_2)}(\textbf{p}) =
\sum_{\alpha_1'\alpha_2'} \int \frac{d^3\textbf{p}'}{(2\pi)^3}
V_{eff}^{(\alpha_1 \alpha_2 \alpha_1' \alpha_2')} (\textbf{p},
\textbf{p}') \chi_M^{(\alpha_1'\alpha_2')}(\textbf{p}')
\end{eqnarray} where

\begin{eqnarray}
\label{eq14}
V_{eff}^{(\alpha_1 \alpha_2 \alpha_1' \alpha_2')}
(\textbf{p}, \textbf{p}') =
N_{12}^{(\alpha_1\alpha_2)}(\textbf{p}) B^{(\alpha_1 \alpha_2
\alpha_1' \alpha_2')}(\textbf{p},
\textbf{p}')N_{12}^{(\alpha_1'\alpha_2')}(\textbf{p}')
\end{eqnarray}

\begin{eqnarray}
\label{eq15a}
\begin{array}{r r}
B^{(\alpha_1 \alpha_2 \alpha_1' \alpha_2')}(\textbf{p},
\textbf{p}') = \left[ 1 - \frac{\alpha_1\alpha_2
\alpha_1'\alpha_2' (\vec{\sigma}_1\vec{p}) (\vec{\sigma}_2\vec{p})
(\vec{\sigma}_1\vec{p}\,') (\vec{\sigma}_2\vec{p}\,')}{
(\omega_1+\alpha_1 m_1) (\omega_2+\alpha_2 m_2)
(\omega_1'+\alpha_1' m_1) (\omega_2'+\alpha_2' m_2) } \right]
V_1(\vec{p}, \vec{p}\,') & (\mbox{SAL})
\end{array}
\\ \nonumber
\\
\label{eq15b}
\begin{array}{r r}
B^{(\alpha_1 \alpha_2 \alpha_1' \alpha_2')}(\textbf{p},
\textbf{p}') = \left[ V_1(\vec{p}, \vec{p}\,') + \frac{\alpha_2
\alpha_2' (\vec{\sigma}_2\vec{p}) (\vec{\sigma}_2\vec{p}\,')}{
(\omega_2+\alpha_2 m_2) (\omega_2'+\alpha_2' m_2) }  V_2(x
;\vec{p}, \vec{p}\,')\right] & (\mbox{GR})
\end{array}
\\ \nonumber
\\
\label{eq15c}
\begin{array}{r r}
B^{(\alpha_1 \alpha_2 \alpha_1' \alpha_2')}(\textbf{p},
\textbf{p}') = \left[ 1 - \frac{\alpha_1\alpha_2
\alpha_1'\alpha_2' (\vec{\sigma}_1\vec{p}) (\vec{\sigma}_2\vec{p})
(\vec{\sigma}_1\vec{p}\,') (\vec{\sigma}_2\vec{p}\,')}{
(\omega_1+\alpha_1 m_1) (\omega_2+\alpha_2 m_2)
(\omega_1'+\alpha_1' m_1) (\omega_2'+\alpha_2' m_2) } \right]
V_1(\vec{p}, \vec{p}\,') &
\\
\\
+\left[ \frac{\alpha_1 \alpha_1' (\vec{\sigma}_1\vec{p})
(\vec{\sigma}_1\vec{p}\,')}{ (\omega_1+\alpha_1 m_1)
(\omega_1'+\alpha_1' m_1) } + \frac{\alpha_2 \alpha_2'
(\vec{\sigma}_2\vec{p}) (\vec{\sigma}_2\vec{p}\,')}{
(\omega_2+\alpha_2 m_2) (\omega_2'+\alpha_2' m_2) } \right] V_2(x;
\vec{p},\vec{p}\,' ) & (\mbox{MW})
\end{array}
\\ \nonumber
\\
\label{eq15d}
\begin{array}{r r}
B^{(\alpha_1 \alpha_2 \alpha_1' \alpha_2')}(\textbf{p},
\textbf{p}') =  V_1(\vec{p}, \vec{p}\,') + \left[ \frac{\alpha_1
\alpha_1' (\vec{\sigma}_1\vec{p}) (\vec{\sigma}_1\vec{p}\,')}{
(\omega_1+\alpha_1 m_1) (\omega_1'+\alpha_1' m_1) } \mu_2 +
\frac{\alpha_2 \alpha_2' (\vec{\sigma}_2\vec{p})
(\vec{\sigma}_2\vec{p}\,')}{ (\omega_2+\alpha_2 m_2)
(\omega_2'+\alpha_2' m_2) } \mu_1 \right] V_2(x ;\vec{p}, \vec{p}\,') & \\
(\mbox{CJ, MNK})
\end{array}
\end{eqnarray}

\begin{eqnarray}
\label{eq16} \omega_i' = \sqrt{m_i^2+{\mathbf{p'}}^2}, \;
V_1(\textbf{p},\textbf{p}') = V_{OG}(\textbf{p},\textbf{p}') +
V_C(\textbf{p},\textbf{p}'), \nonumber
\\
V_2(x;\textbf{p},\textbf{p}') = V_{OG}(\textbf{p},\textbf{p}') +
(2x-1) V_C(\textbf{p},\textbf{p}').
\end{eqnarray}

It is very important that the two-body Dirac equation (\ref{eq13})
with effective potential (\ref{eq14}) with (\ref{eq15a}) obtained
from the equation (\ref{eq1}), corresponding to SAL version
(\ref{eq2}) does not depend on parameter $x$ interned in the
interaction operator (\ref{eq10}), which means that from this
equation can not be obtained any information on the Lorentz (spin)
structure of the confining $q\bar{q}$ interaction potential
(\ref{eq10}). Second interesting result is that the wave functions
satisfying the two-body Dirac equation (\ref{eq13}) with effective
potentials (\ref{eq14}) with expression (\ref{eq15a}, \ref{eq15b})
obtained for SAL and GR versions (\ref{eq2}) of the
3D-relativistic equations have all nonzero "frequency components"
whereas two components of the wave functions satisfying the
equation (\ref{eq1}) with projection operators (\ref{eq2}), are
zero, namely:

\begin{eqnarray}
\label{eq17}
\begin{array}{r r}
\tilde{\Phi}^{(\pm\mp)}_M = \Lambda^{(\pm\mp)}_{12} \tilde{\Phi}_M
= 0 & (\mbox{SAL}) \\
\tilde{\Phi}^{(-\pm)}_M = \Lambda^{(-\pm)}_{12} \tilde{\Phi}_M = 0
& (\mbox{GR})
\end{array}
\end{eqnarray} which directly follows (and is well known) from the
eq.(\ref{eq1}) if it is multiplied (from left) by the operators
$\Lambda^{(\pm\mp)}_{12}$, $\Lambda^{(-\pm)}_{12}$ and used the
formulae (\ref{eq3}).

For formulation normalization condition for the wave function
(\ref{eq8}) which satisfies the equation (\ref{eq7}), we note that
normalization condition for Salpeter wave function obtained in
Ref.\cite{b1} (see relation (3.14)) can be written in the form

\begin{eqnarray}
\label{eq18a}
<\tilde{\Phi}_M |\Pi^{SAL}|\tilde{\Phi}_M > = 2M
\end{eqnarray}

The analogous condition can be derived for wave function
satisfying Gross equation (\ref{eq1}, \ref{eq2}) if we use
equation for full Green operator corresponding to the equation
(\ref{eq1})

\begin{eqnarray}
\label{eq19} \tilde{G}= g_0 \Pi\Gamma_0 + g_0 \tilde{U}
\tilde{G},\; g_0 = \left[ M - h_1 - h_2  \right]^{-1}, \; \Gamma_0
= \gamma_1^0\otimes\gamma_2^0,\; \tilde{U}=\Pi\Gamma_0V
\end{eqnarray} Assuming that the operator $\tilde{G}^{-1}$ exists
(being natural at any rate in the bound states, we need) from
eq.(\ref{eq19}) after some transformations can be obtained the
following relation

\begin{eqnarray}
\label{eq20} \tilde{G}\Gamma_0\Pi \left[ g_0^{-1} - \tilde{U}
\right] \tilde{G}\Gamma_0 = \tilde{G}\Gamma_0\Pi\Pi.
\end{eqnarray}
Noting that $\Pi_{GR}\Pi_{GR} = \Pi_{GR}$ from (\ref{eq20}) we
have

\begin{eqnarray}
\label{eq21} \tilde{G}\Gamma_0\Pi^{GR}\left[ g_0^{-1} - \tilde{U}
\right]\tilde{G}\Gamma_0\Pi^{GR} = \tilde{G}\Gamma_0\Pi^{GR}.
\end{eqnarray}
Now using the spectral representation of Green operator
$\tilde{G}$

\begin{eqnarray}
\label{eq22} \tilde{G}(\textbf{p}) = \sum_B
\frac{|\tilde{\Phi}_{P_B}><\tilde{\bar{\Phi}}_{P_B}|}{P^2-M_B^2} +
\tilde{R}(P), \;\; <\tilde{\bar{\Phi}}_{P_B}| =
<\tilde{\Phi}_{P_B}|\Gamma_0
\end{eqnarray}
from (\ref{eq21}) can be obtained the relation

\begin{eqnarray}
\label{eq23} <\tilde{\Phi}_M |\Pi^{GR}|\tilde{\Phi}_M >\Pi^{GR} =
2M\Pi^{GR}
\end{eqnarray}
It means that the normalization condition analogous to
(\ref{eq18a})

\begin{eqnarray}
\label{eq18b} <\tilde{\Phi}_M |\Pi^{GR}|\tilde{\Phi}_M > = 2M
\end{eqnarray}
holds only in corresponding subspace of the Gilbert space. Note
that the condition (\ref{eq18b}) can be obtained from the formula
(3.28) of ref.\cite{b1}, which was not derived, but supposed with
an analogy to (3.14).

Now, noting that the two-body Dirac equations (\ref{eq7}) for the
SAL and GR versions of the 3D-relativistic equation (\ref{eq1})
were obtained in the approximation (\ref{eq4}) for the projection
operators $\Pi^{SAL}$ and $\Pi^{GR}$, the corresponding condition
for wave function can be obtained from (\ref{eq18a}, \ref{eq18b})
by replacement $\Pi^{SAL} \Rightarrow \Pi^{SAL}_0$ (\ref{eq5b})
and $\Pi^{GR} \Rightarrow \Pi^{GR}_0$ (\ref{eq6}). Thus we have

\begin{eqnarray}
\label{eq24}  <\Psi_M |\Pi^{SAL, GR}_0|\Psi_M
> = 2M
\end{eqnarray}where $\Pi^{SAL}_0$ and $\Pi^{GR}_0$ are given by formulae (\ref{eq5b}), (\ref{eq6}).

Further, noting that the projection operator $\Pi^{MW}_0$
(\ref{eq6b}) satisfies condition $\Pi^{MW}_0 \Pi^{MW}_0 = 1$, from
relation (\ref{eq20}) can be obtained the normalization condition
analogous to (\ref{eq24}) i.e.

\begin{eqnarray}
\label{eq25}  <\Psi_M |\Pi^{MW}_0|\Psi_M
> = 2M.
\end{eqnarray}

As to normalization conditions for wave functions satisfying the
two-body Dirac equation (\ref{eq7}), corresponding to the CJ and
MNK versions, they can not be derived analogously because the
corresponding projection operators $\Pi_0$ (\ref{eq6c})does not
satisfy the conditions $\Pi_0\Pi_0=\Pi_0$ or $\Pi_0\Pi_0=1$. But
bellow we assume (suppose) that the condition analogous to
(\ref{eq25}) can be written in common form

\begin{eqnarray}
\label{eq26}  <\Psi_M |\Pi_0|\Psi_M> = 2M.
\end{eqnarray}
where operator $\Pi_0$ is given by the formulae (\ref{eq5b},
\ref{eq6}, \ref{eq6b}, \ref{eq6c}) for all versions. As a result
with an account of the formulae (\ref{eq8}, \ref{eq11},
\ref{eq12}) the normalization condition for the components of the
wave functions $\chi_M^{(\alpha_1 \alpha_2)}$ takes the form

\begin{eqnarray}
\label{eq27} \sum_{\alpha_1\alpha_2\beta_1\beta_2}
<\chi_M^{(\alpha_1\alpha_2)}|N_{12}^{(\alpha_1\alpha_2)}
f_{12}^{(\alpha_1\alpha_2)+}\Pi_0 f_{12}^{(\beta_1\beta_2)}
N_{12}^{(\beta_1\beta_2)} | \chi_M^{(\beta_1\beta_2)} > = 2M
\end{eqnarray} from which follows

\begin{eqnarray}
\label{eq28} \int \frac{d^3\textbf{p}}{(2\pi)^3}
\sum_{\alpha_1\alpha_2\beta_1\beta_2} \frac{1}{4} \biggl[
E_{12}^{(\alpha_1\alpha_2\beta_1\beta_2)} \left(
\begin{array}{l}
1 \\ 1 \\ 1 \\ 1
\end{array} \right) + \alpha_1\beta_1
E_{12}^{(-\alpha_1\alpha_2-\beta_1\beta_2)}\left(\begin{array}{l} 0 \\
0
\\ 1 \\ \mu_2
\end{array}\right)
+ \alpha_2\beta_2
E_{12}^{(\alpha_1-\alpha_2\beta_1-\beta_2)}\left(\begin{array}{l}
0 \\ 1
\\ 1 \\ \mu_1
\end{array}\right)-
\nonumber \\
-\alpha_1\alpha_2\beta_1\beta_2
E_{12}^{(-\alpha_1-\alpha_2-\beta_1-\beta_2)}\left(\begin{array}{l} 1 \\
0
\\ 1 \\ 0
\end{array}\right) \biggl] \chi_M^{(\alpha_1\alpha_2)*}(\textbf{p})
\chi_M^{(\beta_1\beta_2)}(\textbf{p})= 2M, \;\;
\left( \begin{array}{c} \mbox{SAL} \\ \mbox{GR} \\ \mbox{MW} \\
\mbox{CJ, MNK}
\end{array} \right)
\end{eqnarray}
where

\begin{eqnarray}
\label{eq28e} E_{12}^{(\alpha_1\alpha_2\beta_1\beta_2)} =
\sqrt{(1+\alpha_1\frac{m_1}{\omega_1})(1+\alpha_2\frac{m_2}{\omega_2})
(1+\beta_1\frac{m_1}{\omega_1})(1+\beta_2\frac{m_2}{\omega_2}) }
\end{eqnarray}
Now we use the partial-wave expansion for the function
$\chi_M^{(\alpha_1\alpha_2)}(\textbf{p})$ \cite{b1}

\begin{eqnarray}
\label{eq29} \chi_M^{(\alpha_1\alpha_2)}(\textbf{p}) =
\sum_{LSJM_J}
<\hat{\textbf{n}}|LSJM_J>R_{LSJ}^{(\alpha_1\alpha_2)}(p) \equiv
\sum_{JM_J}\chi_{JM_J}^{(\alpha_1\alpha_2)}(\textbf{p}), \;\;\;
(\textbf{n}=\frac{\textbf{p}}{p})
\end{eqnarray}
where $R_{LSJ}^{(\alpha_1\alpha_2)}(p)$ are corresponding radial
wave functions. And the potential functions
$V_{OG}(\textbf{p},\textbf{p}')$, $V_C(\textbf{p},\textbf{p}')$
are represented in form (local potentials)

\begin{eqnarray}
\label{eq30} V(\textbf{p}-\textbf{p}') = (2\pi)^3
\sum_{\bar{L}\bar{S}\bar{J}\bar{M}_J}
V^{\bar{L}}(\textbf{p},\textbf{p}')
<\textbf{n}|\bar{L}\bar{S}\bar{J}\bar{M}_J>
<\bar{L}\bar{S}\bar{J}\bar{M}_J|\textbf{n}'>
\end{eqnarray}
where

\begin{eqnarray}
\label{eq31} V^{\bar{L}}(\textbf{p},\textbf{p}') =
\frac{2}{\pi}\int_{0}^{\infty}
j_{\bar{L}}(pr)V(r)j_{\bar{L}}(p'r)r^2dr
\end{eqnarray}
$j_{\bar{L}}(x)$ being the spherical Bessel function. Then from
the system of equations (\ref{eq13}), the effective potentials of
which is defined by the formulae (\ref{eq14}-\ref{eq16}) we obtain
the following system of equations for the radial functions
$R_{LSJ}^{(\alpha_1\alpha_2)}(p)$

\textbf{SAL version}
\begin{eqnarray}
\label{eq32a} \left[ M - (\alpha_1\omega_1+\alpha_2\omega_2)
\right]R_{J(_1^0)J}^{(\alpha_1 \alpha_2)}(p) =
\sum_{\alpha_1'\alpha_2'} \int_{0}^{\infty}p'^2dp'
\biggl[\biggl(N_{12}^{(\alpha_1\alpha_2)}(p)N_{12}^{(\alpha_1'\alpha_2')}(p')-
\nonumber \\
- \alpha_1\alpha_2\alpha_1'\alpha_2'
N_{12}^{(-\alpha_1-\alpha_2)}(p)N_{12}^{(-\alpha_1'-\alpha_2')}(p')\biggl)
V_1^J(p,p')\biggl]R_{J(_1^0)J}^{(\alpha_1' \alpha_2')}(p')
\\
\label{eq32b} \left[ M - (\alpha_1\omega_1+\alpha_2\omega_2)
\right]R_{J\pm11J}^{(\alpha_1 \alpha_2)}(p) =
\sum_{\alpha_1'\alpha_2'} \int_{0}^{\infty}p'^2dp' \biggl\{
\biggl[
N_{12}^{(\alpha_1\alpha_2)}(p)N_{12}^{(\alpha_1'\alpha_2')}(p')V_1^{J\pm1}(p,p')-
\nonumber \\
- \alpha_1\alpha_2\alpha_1'\alpha_2'
N_{12}^{(-\alpha_1-\alpha_2)}(p)N_{12}^{(-\alpha_1'-\alpha_2')}(p')
V_{1(J\pm1)}(p,p')\biggl]R_{J\pm11J}^{(\alpha_1' \alpha_2')}(p')-
\nonumber
\\
- \biggl[ \alpha_1\alpha_2\alpha_1'\alpha_2'
N_{12}^{(-\alpha_1-\alpha_2)}(p)N_{12}^{(-\alpha_1'-\alpha_2')}(p')
\frac{2}{2J+1}V_{1 \ominus J}(p,p') \biggl]
R_{J\mp11J}^{(\alpha_1' \alpha_2')}(p') \biggl\}
\end{eqnarray}

\textbf{GR version}

\begin{eqnarray}
\label{eq33a} \left[ M - (\alpha_1\omega_1+\alpha_2\omega_2)
\right]R_{J(_1^0)J}^{(\alpha_1 \alpha_2)}(p) =
\sum_{\alpha_1'\alpha_2'} \int_{0}^{\infty}p'^2dp'\biggl\{ \biggl[
N_{12}^{(\alpha_1\alpha_2)}(p)N_{12}^{(\alpha_1'\alpha_2')}(p')V_1^J(p,p')+
\nonumber \\
+\alpha_2\alpha_2'N_{12}^{(\alpha_1-\alpha_2)}(p)N_{12}^{(\alpha_1'-\alpha_2')}(p')
V_{2\oplus J}^{(^0_1)}(x;p,p') \biggl]R_{J(_1^0)J}^{(\alpha_1'
\alpha_2')}(p')- \nonumber
\\
-\biggl[\alpha_2\alpha_2'N_{12}^{(\alpha_1-\alpha_2)}(p)N_{12}^{(\alpha_1'-\alpha_2')}(p')
V_{2 \ominus J}(x;p,p')\biggl]R_{J(_0^1)J}^{(\alpha_1'
\alpha_2')}(p')
\\
\label{eq33b} \left[ M - (\alpha_1\omega_1+\alpha_2\omega_2)
\right]R_{J\pm11J}^{(\alpha_1 \alpha_2)}(p) =
\sum_{\alpha1'\alpha_2'} \int_{0}^{\infty}p'^2dp' \biggl[
N_{12}^{(\alpha_1\alpha_2)}(p)N_{12}^{(\alpha_1'\alpha_2')}(p')V_1^{J\pm1}(p,p')+
\nonumber
\\
+\alpha_2\alpha_2'N_{12}^{(\alpha_1-\alpha_2)}(p)N_{12}^{(\alpha_1'-\alpha_2')}(p')
V_2^{J}(x;p,p') \biggl]R_{J\pm11J}^{(\alpha_1' \alpha_2')}(p')
\end{eqnarray}

\textbf{MW, CJ and MNK versions}

\begin{eqnarray}
\label{eq34a} \left[ M - (\alpha_1\omega_1+\alpha_2\omega_2)
\right]R_{J(_1^0)J}^{(\alpha_1 \alpha_2)}(p) =
\sum_{\alpha1'\alpha_2'} \int_{0}^{\infty}p'^2dp' \biggl\{ \biggl[
\biggl(
N_{12}^{(\alpha_1\alpha_2)}(p)N_{12}^{(\alpha_1'\alpha_2')}(p')
\left(
\begin{array}{l}
1 \\ 1
\end{array}\right)-
\nonumber
\\ -\alpha_1 \alpha_2\alpha_1' \alpha_2' N_{12}^{(-\alpha_1-\alpha_2)}(p)N_{12}^{(-\alpha_1'-\alpha_2')}(p')
\left(
\begin{array}{l}
1 \\ 0
\end{array}\right) \biggl)V_1^J(p,p')+
\nonumber
\\
+\biggl(\alpha_1\alpha_1'N_{12}^{(-\alpha_1\alpha_2)}(p)N_{12}^{(-\alpha_1'\alpha_2')}(p')
\left(
\begin{array}{l}
1 \\ \mu_2
\end{array}\right)+
\nonumber
\\
+\alpha_2\alpha_2'N_{12}^{(\alpha_1-\alpha_2)}(p)N_{12}^{(\alpha_1'-\alpha_2')}(p')
\left(
\begin{array}{l}
1 \\ \mu_1
\end{array}\right) \biggl) V_{2\oplus J}^{(^0_1)}(x;p,p')\biggl]R_{J(_1^0)J}^{(\alpha_1'
\alpha_2')}(p')+
\nonumber
\\
+\biggl[ \biggl(
\alpha_1\alpha_1'N_{12}^{(-\alpha_1\alpha_2)}(p)N_{12}^{(-\alpha_1'\alpha_2')}(p')
\left(
\begin{array}{l}
1 \\ \mu_2
\end{array}\right)
-\alpha_2\alpha_2'N_{12}^{(\alpha_1-\alpha_2)}(p)N_{12}^{(\alpha_1'-\alpha_2')}(p')
\left(
\begin{array}{l}
1 \\ \mu_1
\end{array}\right)
\biggl)\times \nonumber
\\
\times V_{2 \ominus J}(x;p,p') \biggl] R_{J(_0^1)J}^{(\alpha_1'
\alpha_2')}(p')\biggl\} \;\;\; \left(
\begin{array}{c}
\mbox{MW} \\ \mbox{CJ,MNK}
\end{array}\right)
\end{eqnarray}

\begin{eqnarray}
\label{eq34b} \left[ M - (\alpha_1\omega_1+\alpha_2\omega_2)
\right]R_{J\pm11J}^{(\alpha_1 \alpha_2)}(p) =
\sum_{\alpha1'\alpha_2'} \int_{0}^{\infty}p'^2dp' \biggl\{ \biggl[
N_{12}^{(\alpha_1\alpha_2)}(p)N_{12}^{(\alpha_1'\alpha_2')}(p')
\left(
\begin{array}{l}
1 \\ 1
\end{array}\right) V_1^{J\pm1}(p,p') -
\nonumber
\\ -\alpha_1 \alpha_2\alpha_1' \alpha_2' N_{12}^{(-\alpha_1-\alpha_2)}(p)N_{12}^{(-\alpha_1'-\alpha_2')}(p')
\left(
\begin{array}{l}
1 \\ 0
\end{array}\right) V_{1(J\pm1)}(p,p')+
\nonumber
\\
+\biggl(\alpha_1\alpha_1'N_{12}^{(-\alpha_1\alpha_2)}(p)N_{12}^{(-\alpha_1'\alpha_2')}(p')
\left(
\begin{array}{l}
1 \\ \mu_2
\end{array}\right)+
\nonumber
\\
+\alpha_2\alpha_2'N_{12}^{(\alpha_1-\alpha_2)}(p)N_{12}^{(\alpha_1'-\alpha_2')}(p')
\left(
\begin{array}{l}
1 \\ \mu_1
\end{array}\right) \biggl) V_2^J(x;p,p')\biggl]R_{J\pm11J}^{(\alpha_1'
\alpha_2')}(p')+
\nonumber
\\
+\biggl[
\alpha_1\alpha_1'\alpha_2\alpha_2'N_{12}^{(-\alpha_1-\alpha_2)}(p)N_{12}^{(-\alpha_1'-\alpha_2')}(p')
\frac{2}{2J+1} V_{1\ominus J} (p,p') \biggl]
R_{J\mp11J}^{(\alpha_1' \alpha_2')}(p')\biggl\}
\;\;\; \left(
\begin{array}{c}
\mbox{MW} \\ \mbox{CJ,MNK}
\end{array}\right)
\end{eqnarray}
where

\begin{eqnarray}
V_{n \oplus J}^{(^0_1)} = \frac{1}{2J+1} \biggl[ (^{J+1}_{
J})V_n^{J+1} + (^{ J}_{J+1})V_n^{J-1} \biggl] \nonumber
\\
V_{n \ominus J} = \frac{\sqrt{J(J+1)}}{2J+1}
\biggl[V_n^{J+1}-V_n^{J-1} \biggl], \;\;\;\; n = 1,2
\\
V_{n (J\pm1)} =\frac{1}{(2J+1)^2} \biggl[ V_n^{J\pm1} + 4J(J+1)
V_n^{J\mp1} \biggl] \nonumber
\end{eqnarray}

It is interesting to compare the system of equations
(\ref{eq32a}-\ref{eq34b}) with the system of equations obtained
from (\ref{eq1}) without the approximation (\ref{eq4}, \ref{eq4b})
(see eqs. (4.16, 17) in \cite{b1}, neglecting the terms
corresponding to t'Hooft interaction )

\begin{eqnarray}
\label{eq36a} \left[ M - (\alpha_1\omega_1+\alpha_2\omega_2)
\right]R^{(\alpha_1\alpha_2)}_{J(^0_1)J}(p) =
A^{(\alpha_1\alpha_2)}(M;p) \sum_{\alpha_1'\alpha_2'}
\int_0^{\infty} p'^2dp' \biggl\{ \biggl[ \biggl(
N_{12}^{(\alpha_1\alpha_2)}(p)N_{12}^{(\alpha_1'\alpha_2')}(p')+
\nonumber
\\
+\alpha_1\alpha_2\alpha_1'\alpha_2'
N_{12}^{(-\alpha_1-\alpha_2)}(p)N_{12}^{(-\alpha_1'-\alpha_2')}(p')
\biggl) V_1^J(p,p')+ \nonumber
\\
+\biggl( \alpha_1\alpha_1'
N_{12}^{(-\alpha_1\alpha_2)}(p)N_{12}^{(-\alpha_1'\alpha_2')}(p')
+ \alpha_2\alpha_2'
N_{12}^{(\alpha_1-\alpha_2)}(p)N_{12}^{(\alpha_1'-\alpha_2')}(p')
\biggl) V_{2\oplus J}^{(^0_1)}(x;p,p') \biggl]
R^{(\alpha_1'\alpha_2')}_{J(^0_1)J}(p') - \nonumber
\\
- \biggl[ \biggl( \alpha_1\alpha_1'
N_{12}^{(-\alpha_1\alpha_2)}(p)N_{12}^{(-\alpha_1'\alpha_2')}(p')
- \alpha_2\alpha_2'
N_{12}^{(\alpha_1-\alpha_2)}(p)N_{12}^{(\alpha_1'-\alpha_2')}(p')
\biggl) V_{2\ominus J}(x;p,p')
\biggl]R^{(\alpha_1'\alpha_2')}_{J(^1_0)J}(p') \biggl\}
\end{eqnarray}

\begin{eqnarray}
\label{eq36b} \left[ M - (\alpha_1\omega_1+\alpha_2\omega_2)
\right]R^{(\alpha_1\alpha_2)}_{J\pm11J}(p) =
A^{(\alpha_1\alpha_2)}(M;p) \sum_{\alpha_1'\alpha_2'}
\int_0^{\infty} p'^2dp' \times \nonumber
\\
\times \biggl\{ \biggl[ \biggl(
N_{12}^{(\alpha_1\alpha_2)}(p)N_{12}^{(\alpha_1'\alpha_2')}(p')V_1^{J\pm1}(p,p')+
\nonumber
\\
+\alpha_1\alpha_2\alpha_1'\alpha_2'
N_{12}^{(-\alpha_1-\alpha_2)}(p)N_{12}^{(-\alpha_1'-\alpha_2')}(p')
V_{1(J\pm 1)}(p,p') \biggl)+  \nonumber
\\
+\biggl( \alpha_1\alpha_1'
N_{12}^{(-\alpha_1\alpha_2)}(p)N_{12}^{(-\alpha_1'\alpha_2')}(p')
+ \alpha_2\alpha_2'
N_{12}^{(\alpha_1-\alpha_2)}(p)N_{12}^{(\alpha_1'-\alpha_2')}(p')
\biggl) V_2^J(x;p,p') \biggl]
R^{(\alpha_1'\alpha_2')}_{J\pm11J}(p')+ \nonumber
\\
+ \biggl[ \alpha_1\alpha_2\alpha_1'\alpha_2'
N_{12}^{(-\alpha_1-\alpha_2)}(p)N_{12}^{(-\alpha_1'-\alpha_2')}(p')
\frac{2}{2J+1}V_{1\ominus J}(x;p,p')
\biggl]R^{(\alpha_1'\alpha_2')}_{J\mp11J}(p') \biggl\}
\end{eqnarray} where

\begin{eqnarray}
\label{eq37} A^{(\pm\pm)} = \pm 1, \; A^{(\pm\mp)} = 0, \;
\mbox{(SAL)}; \;\; A^{(+\pm)} = \pm 1, \; A^{(-\pm)} = 0, \;
\mbox{(GR)}; \nonumber
\\
A^{(\pm\pm)} = \pm 1, \; A^{(\pm\mp)} =
\frac{M}{\omega_1+\omega_2}, \; \mbox{(MW)}; \;\;
A^{(\alpha_1\alpha_2)} =
\frac{M+(\alpha_1\omega_1+\alpha_2\omega_2)}{2(\omega_1+\omega_2)},
\; \mbox{(CJ)}; \;\;
\\ A^{(\alpha_1\alpha_2)} =
\frac{1}{2BR}\biggl[ M(a-(p_0^{+})^2) -
p_0^{+}M(\alpha_1\omega_1-\alpha_2\omega_2) +
B(\alpha_1\omega_1+\alpha_2\omega_2) \biggl], \; \mbox{(MNK)}
\nonumber
\end{eqnarray}

Note that the last expression in (\ref{eq37}) is obtained from
(3.62) in \cite{b1} after some transformation.

Main difference between the system of equations
(\ref{eq32a}-\ref{eq34b}) and (\ref{eq36a}, \ref{eq36b})with the
expression (\ref{eq37}) is following: 1) In the wave functions
$R^{(\alpha_1\alpha_2)}_{LSJ}$ satisfying the system of equations
(\ref{eq36a}, \ref{eq36b}) for (SAL) and (GR) versions the nonzero
functions are only $R^{(\pm\pm)}_{LSJ}$ and $R^{(+\pm)}_{LSJ}$
respectively (about this fact was mentioned above), whereas in the
corresponding system of equations (\ref{eq32a}), (\ref{eq32b}),
(\ref{eq33a}), (\ref{eq33b}) all components of wave functions
$R^{(\alpha_1\alpha_2)}_{LSJ}$ are nonzero; 2) The system of
equations (\ref{eq36a}, \ref{eq36b}) for the MW, CJ and MNK
versions are nonlinear over $M$, whereas the system of equations
(\ref{eq34a}, \ref{eq34b}) are linear one;
3) Dirac equations (\ref{eq32a},\ref{eq32b}) 
obtained from the Salpeter, equation do not
depend on $x$.

\section{Procedure for solving the obtained equations}

For solving bound-state equations (\ref{eq32a}-\ref{eq34b}) or
(\ref{eq36a},\ref{eq36b}), we need to specify the interquark
interaction potentials $V_{OG}$ and $V_C$ (\ref{eq10}). Below for
$V_C(r)$ we use the following form \cite{b1}, \cite{b11}

\begin{eqnarray}
\label{eq38a} V_C(r) = \frac{4}{3}\alpha_S(m_{12}^2)\left(
\frac{\mu_{12}\omega_0^2r^2}{2\sqrt{1+A_0m_1m_2r^2}} - V_0 \right)
\end{eqnarray}

\begin{eqnarray}
\label{eq38b} \alpha_S(Q^2) = \frac{12\pi}{33-2n_f}\left[
\ln{\frac{Q^2}{\Lambda^2}} \right]^{-1}, \;\; m_{12}=m_1+m_2, \;\;
\mu_{12}=\frac{m_1m_2}{m_{12}}
\end{eqnarray} where $Q^2$ is the momentum transferred and the
$\frac{4}{3}$comes from the color-dependent part of the $q\bar{q}$
interaction, $n_f$ is the number of flavors ($n_f=3$ for $u,d,s$
quarks; $n_f=4$ for $u,d,s,c$; $n_f=5$ for $u,d,s,c,b$).
$\omega_0$, $A_0$, $V_0$ and $\Lambda$ are considered to be the
free parameters of the model. The potential given by expression
(\ref{eq38a}) effectively reduces to the harmonic oscillator
potential for the light quarks $u,d,s$ and to the linear potential
to the heavy $c,b$ quarks if the dimensionless parameter $A_0$ is
chosen small enough. Moreover, asymptotically, for a large $r$  it
is linear and almost flavor-independent. The one-gluon exchange
potential is given by standard expression \cite{b1}, \cite{b11}

\begin{eqnarray}
\label{eq39} V_{OG}(r) = -\frac{4}{3}\frac{\alpha_S(m_{12}^2)}{r}
\end{eqnarray}

Now we have to specify the numerical procedure for solution of the
systems of radial equations
(\ref{eq32a}-\ref{eq34a}),(\ref{eq36a}),(\ref{eq36b}). A possible
algorithm looks as follows: we choose the known basis functions
denoted by $R_{nL}(p)$. The unknown radial wave functions are
expanded in the linear combination of the basis functions

\begin{eqnarray}
\label{eq40} R_{LSJ}^{(\alpha_1\alpha_2)}(p) = \sqrt{2M(2\pi)^3}
\sum_{n=0}^{\infty}C_{nLSJ}^{(\alpha_1\alpha_2)}R_{nL}(p)
\end{eqnarray} where $C_{nLSJ}^{(\alpha_1\alpha_2)}$ are the
coefficients of the expansion. The integral equation for the
radial wave functions is then transformed into the system of
linear equations for these coefficients. If the transaction is
carried out the finite system of equations is obtained that can be
solved by using conventional numerical methods. The convergence of
the whole procedure, with more terms taken into account in the
expansion (\ref{eq40}) depend on the successful choice of the
basis. In case of the confining potential of form (\ref{eq38a}) it
is natural to take as a basis the functions corresponding to
oscillator potential, which is obtained from (\ref{eq38a}) at
$A_0=0$, in non-relativistic limit of the system of equations
obtained from from
(\ref{eq32a}-\ref{eq34b}),(\ref{eq36a}),(\ref{eq36b}). The radial
wave functions in this case have the form \cite{b1}(the formula
(4.52)).

\begin{eqnarray}
\label{eq41}
\begin{array}{l}
R_{nL}(p) = p_0^{-3/2}R_{nL}(z), \;\;\; p_0 =
\sqrt{\mu_{12}\omega_0\sqrt{{\frac{4}{3}\alpha_S(m_{12}^2)}}},
\;\;\; z = \frac{p}{p_0}
\\
R_{nL}(z) = c_{nL}z^L
\exp(-\frac{z^2}{2})_1F_1(-n,L+\frac{3}{2},z^2), \;\;\; c_{nL} =
\sqrt{\frac{2\Gamma(n+L+\frac{3}{2})}{\Gamma(n+1)}}\frac{1}{\Gamma(L+\frac{3}{2})}
\end{array}
\end{eqnarray} where $_1F_1$ denotes the confluent hypergeometric
function.

Now, satisfying the expression (\ref{eq40}) into the system of
equations (\ref{eq32a}-\ref{eq34b}),(\ref{eq36a}),(\ref{eq36b}),
the following algebraic equations for the coefficients
$C_{nLSJ}^{(\alpha_1\alpha_2)}$ can be obtained

\begin{eqnarray}
\label{eq42} M C_{nLSJ}^{(\alpha_1\alpha_2)} =
\sum_{\alpha_1'\alpha_2'} \sum_{n'L'S'}
H^{(\alpha_1\alpha_2;\alpha_1',\alpha_2')}_{nLSJ;n'L'S'J}(M)C_{n'L'S'J}^{(\alpha_1'\alpha_2')}
\end{eqnarray}

It is necessary to note that the matrix $H_{\alpha\beta}(M)$
depends on meson mass $M$ only for MW, CJ and MNK versions as it
can be seen from equations (\ref{eq42}) for $M$ is not linear one
and therefore should be solved, e.g. by iteration. As to the
system of Dirac equations (\ref{eq32a}-\ref{eq34b}) such a problem
does not exist.

\begin{table}
\begin{tabular}{| r | c c c c c c c |}
\hline $x$ & $\;\;\;$ 0.0 $\;\;\;$ & $\;\;\;$ 0.1 $\;\;\;$ &
$\;\;\;$ 0.3 $\;\;\;$ & $\;\;\;$ 0.5 $\;\;\;$ & $\;\;\;$ 0.7
$\;\;\;$
& $\;\;\; $0.9 $\;\;\;$ & $\;\;\;$ 1.0 $\;\;\;$ \\
\hline \hline
%
%
& \multicolumn{7}{|c|}{$u\bar{s}$ $^1S_0$ (494)} \\
\hline
SAL  & *   & *   & 853 & 892 & 923 & 948 & 958 \\
SALD & \multicolumn{7}{ c |}{1000} \\
GR   & 819 & 838 & 873 & 906 & *   & *   & *   \\
GRD  & 957 & 956 & 954 & 953 & *   & *   & *   \\
MW   & 847 & 858 & 883 & 909 & *   & *   & *   \\
MWD  & 908 & 917 & 938 & 964 & 998 & *   & *   \\
CJ   & 865 & 878 & 905 & 935 & 968 & *   & *   \\
CJD  & 924 & 930 & 942 & 955 & 972 & *   & *   \\
MNK  & 866 & 793 & 819 & 844 & *   & *   & *   \\
MNKD & 923 & 929 & 941 & 955 & 972 & *   & *   \\
%
%
\hline \hline
& \multicolumn{7}{|c|}{$u\bar{s}$ $^3S_1$ (892)} \\
\hline
SAL  & *   & 812 & 870 & 914 & 950 & 980 & 993 \\
SALD & \multicolumn{7}{ c |}{979} \\
GR   & 839 & 859 & 897 & 934 & 967 & *   & *   \\
GRD  & 944 & 947 & 954 & 962 & 975 & *   & *   \\
MW   & 863 & 877 & 907 & 943 & 983 & *   & *   \\
MWD  & 879 & 887 & 905 & 928 & 957 & *   & *   \\
CJ   & 878 & 893 & 924 & 959 & 998 & *   & *   \\
CJD  & 924 & 930 & 942 & 955 & 972 & *   & *   \\
MNK  & 814 & 830 & 861 & 891 & *   & *   & *   \\
MNKD & 923 & 929 & 941 & 955 & 972 & *   & *   \\
%
%
\hline \hline
& \multicolumn{7}{|c|}{$u\bar{s}$ $^3P_0$ (1350)} \\
\hline
SAL  & 1189 & 1204 & 1213 & 1210 & 1202 & 1189 & 1182 \\
SALD & \multicolumn{7}{ c |}{1349} \\
GR   & 1233 & 1232 & 1229 & 1223 & 1218 & *    & *    \\
GRD  & 1304 & 1302 & 1300 & 1298 & 1298 & *    & *    \\
MW   & 1255 & 1253 & 1249 & 1250 & *    & *    & *    \\
MWD  & 1278 & 1274 & 1267 & 1260 & 1257 & *    & *    \\
CJ   & 1268 & 1267 & 1263 & 1260 & 1264 & *    & *    \\
CJD  & 1296 & 1294 & 1290 & 1287 & 1284 & 1285 & *    \\
MNK  & 1217 & 1212 & 1202 & 1190 & *    & *    & *    \\
MNKD & 1295 & 1293 & 1289 & 1286 & 1284 & *    & *    \\
%
%
\hline \hline
& \multicolumn{7}{|c|}{$u\bar{s}$ $^3P_2$ (1430)} \\
\hline
SAL  & *    & *    & 1189 & 1289 & 1367 & 1430 & 1458 \\
SALD & \multicolumn{7}{ c |}{1318} \\
GR   & 1119 & 1159 & 1237 & 1310 & 1381 & *    & *    \\
GRD  & 1278 & 1284 & 1297 & 1314 & 1336 & *    & *    \\
MW   & 1184 & 1209 & 1262 & 1326 & 1384 & *    & *    \\
MWD  & 1185 & 1200 & 1234 & 1275 & 1326 & *    & *    \\
CJ   & 1181 & 1211 & 1276 & 1345 & 1421 & *    & *    \\
CJD  & 1254 & 1264 & 1286 & 1310 & 1337 & 1369 & 1388 \\
MNK  & 1137 & 1165 & 1223 & 1282 & 1344 & 1408 & *    \\
MNKD & 1254 & 1264 & 1285 & 1309 & 1388 & 1372 & *    \\
\hline
\end{tabular}
\caption{The dependence of the $q\bar{q}$ system mass for light
constituent quarks on the mixing parameter $x$ in the different
3D-reductions of Bethe-Salpeter equations and corresponding Dirac
equations. "*" denotes the absence of the stable solutions. Masses
are given in $MeV$.}
\end{table}


\begin{table}
\begin{tabular}{| r | c c c c c c c |}
\hline $x$ & $\;\;\;$ 0.0 $\;\;\;$ & $\;\;\;$ 0.1 $\;\;\;$ &
$\;\;\;$ 0.3 $\;\;\;$ & $\;\;\;$ 0.5 $\;\;\;$ & $\;\;\;$ 0.7
$\;\;\;$
& $\;\;\; $0.9 $\;\;\;$ & $\;\;\;$ 1.0 $\;\;\;$ \\
\hline
%
%
\hline
& \multicolumn{7}{|c|}{$u\bar{c}$ $^1S_0$ (1863)} \\
\hline
SAL  & 1881 & 1895 & 1920 & 1943 & 1965 & 1985 & 1994 \\
SALD & \multicolumn{7}{ c |}{1983} \\
GR   & 1883 & 1896 & 1921 & 1944 & 1966 & 1986 & 1995 \\
GRD  & 1979 & 1979 & 1978 & 1978 & 1978 & 1978 & 1979 \\
MW   & 1915 & 1922 & 1935 & 1951 & 1972 & *    & *    \\
MWD  & 1924 & 1929 & 1942 & 1958 & 1979 & *    & *    \\
CJ   & 1921 & 1928 & 1943 & 1960 & 1982 & *    & *    \\
CJD  & 1932 & 1937 & 1948 & 1961 & 1978 & 2003 & *    \\
MNK  & 1928 & 1934 & 1946 & 1961 & 1978 & *    & *    \\
MNKD & 1927 & 1932 & 1944 & 1958 & 1977 & *    & *    \\
\hline
%
%
\hline
& \multicolumn{7}{|c|}{$u\bar{c}$ $^3S_1$ (2010)} \\
\hline
SAL  & 1883 & 1897 & 1922 & 1946 & 1968 & 1988 & 1998 \\
SALD & \multicolumn{7}{ c |}{1981} \\
GR   & 1886 & 1899 & 1924 & 1947 & 1969 & 1989 & 1999 \\
GRD  & 1977 & 1977 & 1978 & 1979 & 1981 & 1982 & 1983 \\
MW   & 1918 & 1924 & 1938 & 1955 & 1977 & *    & *    \\
MWD  & 1921 & 1926 & 1939 & 1955 & 1975 & *    & *    \\
CJ   & 1923 & 1930 & 1948 & 1963 & 1981 & *    & *    \\
CJD  & 1932 & 1937 & 1948 & 1961 & 1978 & 2003 & *    \\
MNK  & 1930 & 1935 & 1948 & 1963 & 1981 & *    & *    \\
MNKD & 1927 & 1932 & 1944 & 1958 & 1977 & *    & *    \\
\hline
%
%
\hline
& \multicolumn{7}{|c|}{$s\bar{c}$ $^1S_0$ (1971)} \\
\hline
SAL  & 2020 & 2031 & 2055 & 2070 & 2088 & 2105 & 2113 \\
SALD & \multicolumn{7}{ c |}{2106} \\
GR   & 2023 & 2033 & 2052 & 2071 & 2089 & 2106 & 2114 \\
GRD  & 2106 & 2100 & 2100 & 2100 & 2100 & 2100 & 2100 \\
MW   & 2044 & 2050 & 2062 & 2077 & 2094 & 2118 & *    \\
MWD  & 2052 & 2058 & 2070 & 2084 & 2101 & 2126 & *    \\
CJ   & 2051 & 2057 & 2071 & 2087 & 2105 & 2126 & *    \\
CJD  & 2063 & 2067 & 2077 & 2087 & 2100 & 2116 & 2127 \\
MNK  & 2052 & 2057 & 2069 & 2082 & 2097 & 2116 & *    \\
MNKD & 2059 & 2063 & 2073 & 2085 & 2100 & 2120 & *    \\
\hline
%
%
\hline
& \multicolumn{7}{|c|}{$s\bar{c}$ $^3S_1$ (2107)} \\
\hline
SAL  & 2023 & 2033 & 2054 & 2073 & 2091 & 2108 & 2116 \\
SALD & \multicolumn{7}{ c |}{2104} \\
GR   & 2025 & 2035 & 2055 & 2074 & 2092 & 2110 & 2118 \\
GRD  & 2098 & 2099 & 2100 & 2102 & 2103 & 2105 & 2106 \\
MW   & 2047 & 2053 & 2065 & 2080 & 2098 & 2124 & *    \\
MWD  & 2049 & 2054 & 2066 & 2080 & 2097 & 2121 & *    \\
CJ   & 2053 & 2060 & 2074 & 2089 & 2108 & *    & *    \\
CJD  & 2063 & 2067 & 2077 & 2087 & 2100 & 2116 & 2127 \\
MNK  & 2054 & 2059 & 2071 & 2084 & 2100 & 2119 & *    \\
MNKD & 2059 & 2063 & 2073 & 2085 & 2100 & 2120 & *    \\
\hline
\end{tabular}
\caption{The dependence of the $q\bar{q}$ system mass for heavy
constituent quarks on the mixing parameter $x$ in the different
3D-reductions of Bethe-Salpeter equations and corresponding Dirac
equations. "*" denotes the absence of the stable solutions. Masses
are given in $MeV$.}
\end{table}

\section{The numerical results and discussions}

The main problem we have investigated at first stage is dependence
of the existence of stable solutions of the eq. (\ref{eq42}) i.e.
the equations (\ref{eq32a}-\ref{eq34b}),(\ref{eq36a}),
(\ref{eq36b}) on Lorentz (spin) structure of the confining
$q\bar{q}$ interaction potential, i.e. on the parameter $x$. This
will be done taking as examples the $u\bar{s}$, $c\bar{u}$ and
$c\bar{s}$, bound states with constituent quark masses $m_u = m_d
= 280 \mbox{MeV} $, $m_s = 400 \mbox{MeV} $, $m_c = 1470
\mbox{MeV} $ and the free parameters of the confining potential
(\ref{eq38a},\ref{eq38b}) - $\omega_0 = 710 \mbox{MeV}$, $V_0 =
525 \mbox{MeV}$, $A_0 = 0.0270$, $\Lambda = 120 \mbox{MeV}$.

Note, that in \cite{b11} only the SAL version of 3D-reduction of
Bethe-Salpeter equation was considered as to MW, CJ and MNK
without additional approximation (\ref{eq4}) with oscillator like
potential ($A_0 = 0$ in (\ref{eq38a})) were considered in Refs.
\cite{b7}, \cite{b12}.

The results of the calculations are given for states $^{2S+1}L_J$
(note, that for cases $^3S_1$, $^3P_2$, $^3P_1$ are neglected
additional corresponding terms $^3D_1$, $^3F_2$, $^1P_1$, because
they give small contribution in the calculated mass).

The additional conclusions to pure theoretical
results formulated at the end  of section 2, which follow 
from the tables, are the following:

\begin{itemize}
  \item{The area of dependence on parameter $x$ existence of stable solutions of corresponding
  equations is a little extended for corresponding Dirac equation.}
  \item{The results for CJ and MNK versions for the corresponding Dirac equations are almost
  the same which can be seen from formulae (\ref{eq6c}). Further, it can be shown exactly,
  that Dirac equations in the CJ and MNK versions are equivalent for the states $^1S_0$
  and $^3S_1$ when mixture of states $^3S_1$ and $^3D_1$ is neglected}.
  \item{For $u\bar{c}$} and $s\bar{c}$ bound systems Gross version
  works better what is related to the large difference of
  constituent masses.
  \item{The area of the existence of the stable solutions is enlarged with increasing of the constituent
  masses which is theoretically understandable.}
  \item{Masses of the bound $q\bar{q}$ systems obtained from solutions of Dirac equations are bigger
  then masses corresponding to 3D-equations obtained from BS equation for all versions except GR version
  case.}
\item[]
Note, that for $x=0.5$ the stable solutions always exist.
\end{itemize}

{\em Acknowledgements.} The authors thank A.Rusetsky for useful 
discussions and comments.

%
%


\begin{thebibliography}{99}
\bibitem{b1} T.Kopaleishvili, Phys. Part. Nucl. $\mathbf{32}$ (2001) 560-588
\bibitem{b2} E.E.Salpeter, Phys. Rev. $\mathbf{87}$ (1952) 328
\bibitem{b3} F.Gross, Phys. Rev. $\mathbf{186}$ (1969) 1448;
$\mathbf{C26}$ (1982) 2203, 2226
\bibitem{b4} V.B.Mandelzweig, S.J.Wallace, Phys. Lett.
$\mathbf{B197}$(1987) 469
\bibitem{b5} E.D.Cooper, B.K.Jennings, Nucl.Phys. $\mathbf{A500}$
(1989), 551
\bibitem{b6} K.M.Moung, J.W.Norbury, D.E.Kahana, J.Phys. G.
Nucl. Part. Phys. $\mathbf{22}$ (1996) 315
\bibitem{b7} T.Babutsidze, T.Kopaleishvili, A.Rusetsky, Phys.
Lett. $\mathbf{B426}$ (1998) 139
\bibitem{b8} Z.K.Silagadze, A.A.Khelashvili, Theor.Math.Phys.
$\mathbf{61}$ (1984) 431
\bibitem{b9} C.Semay, R.Cenleneer, Phys. Rev. $\mathbf{D48}$
(1993)4361
\bibitem{b10} H.W.Crater, P.Van Alstine invited paper presented
at a conference on September 12th, 1997 at the Univercity of
Georgia in honour of Professor Donald Robson on his 60th birthday.
\bibitem{b11} A.Archvadze, M.Chachkhunashvili, T.Kopaleishvili,
A.Rusetsky, Nucl. Phys. $\mathbf{A581}$ (1995) 460
\bibitem{b12} T.Babutsidze, T.Kopaleishvili, A.Rusetsky Phys. Rev.
$\mathbf{C59}$ (1999) 976
\end{thebibliography}
\end{document}